\documentclass[graybox, envcountchap]{svmult}

\usepackage{mathptmx}        % selects Times Roman as basic font
\usepackage{helvet}          % selects Helvetica as sans-serif font
\usepackage{courier}         % selects Courier as typewriter font
\usepackage{makeidx}         % allows index generation
\usepackage{graphicx}        % standard LaTeX graphics tool
\usepackage{multicol}        % used for the two-column index
\usepackage[bottom]{footmisc}% places footnotes at page bottom

\makeindex             % used for the subject index
\usepackage{amsmath}
\usepackage{amssymb}
\usepackage{lscape}
\usepackage{multirow}

\def\gapp{\ifmmode\stackrel{>}{_{\sim}}\else$\stackrel{<}{_{\sim}}$\fi}
\def\gsim{\lower.5ex\hbox{\gtsima}}
\def\gtsima{$\; \buildrel > \over \sim \;$}
\def\lapp{\ifmmode\stackrel{<}{_{\sim}}\else$\stackrel{<}{_{\sim}}$\fi}
\def\lsim{\lower.5ex\hbox{\ltsima}}
\def\ltsima{$\; \buildrel < \over \sim \;$}

\newcommand\apgt{\ {\raise-.5ex\hbox{$\buildrel>\over\sim$}}\ }
\newcommand\aplt{\ {\raise-.5ex\hbox{$\buildrel<\over\sim$}}\ }
\begin{document}
\pagestyle{empty}
\frontmatter%%%%%%%%%%%%%%%%%%%%%%%%%%%%%%%%%%%%%%%%%%%%%%%%%%%%%%

\include{dedic}
\include{foreword}
\include{preface}

\mainmatter%%%%%%%%%%%%%%%%%%%%%%%%%%%%%%%%%%%%%%%%%%%%%%%%%%%%%%%

\setcounter{chapter}{10}

\title{The Multiple Origin of Blue Straggler Stars: Theory vs. Observations}

\author{Hagai B. Perets}

\institute{Physics department, Technion - Israel Institute of Technology, Haifa,
Israel 32000,\\
  \email{hperets@physics.technion.ac.il}}

\maketitle
\label{Chapter:Perets}

\abstract*{
In this chapter we review the various suggested channels for the formation
and evolution of blue straggler stars (BSSs) in different environments and
their observational predictions. These include mass transfer during
binary stellar evolution - case A/B/C and D (wind Roche-lobe overflow)
mass transfer, stellar collisions during single and binary encounters
in dense stellar cluster, and coupled dynamical and stellar evolution
of triple systems. We also explore the importance of the BSS and binary
dynamics in stellar clusters. We review the various observed properties
of BSSs in different environments (halo and bulge BSSs, BSSs in globular
clusters and BSSs in old open clusters), and compare the current observations
with the theoretical predictions for BSS formation. We try to constrain
the likely progenitors and processes that play a role in the formation
of BSSs and their evolution. We find that multiple channels of BSS formation
are likely to take part in producing the observed BSSs, and we point
out the strengths and weaknesses of each the formation channel in
respect to the observational constraints. Finally we point out directions
to further explore the origin of BSS, and highlight eclipsing binary
BSSs as important observational tool. 
}

\section{Introduction}
\label{secper:1}

Blue straggler stars (BSSs) are stars that appear to be anomalously young
compared to other stars of their population. In particular, BSSs lie
along an extension of the main sequence\index{main sequence star} (MS) in the colour-magnitude
diagram\index{colour-magnitude
diagram}, a region from which most of the stars of equal mass and age
have already evolved. Such stars appear to be brighter and bluer than
the turn-off\index{turn-off} point of the stellar population in which they were observed.
Their location in the colour-magnitude diagram suggests that BSSs in
old open clusters\index{open cluster} (OCs) and globular clusters\index{globular cluster} (GCs) have typical masses
of $1.2-1.5\, M_{\odot}$, that are significantly larger than those
of normal stars in such systems. Thus, they are thought to have increased
their mass during their evolution. Several mechanisms have been proposed
for their formation: $i)$ stellar collisions\index{collision} due to dynamical interactions
in dense stellar systems \cite{hil+76} $ii)$ coalescence or mass transfer\index{mass transfer}
between two companions due to binary stellar\index{binary star} evolution \cite{mcc64}
$iii)$ induced mergers\index{merger}/collisions\index{collision} through coupled dynamical/stellar
evolution in triple systems\index{triple system} \cite{per+09}. The roles of each of
these mechanisms in producing the observed BSS populations and their
properties are still debated, as each of these scenarios were found
to be successful in explaining some of the BSS observations, but fail
in others.

In this review, we first discuss the observed properties of BSSs in
the different environments (Sect.~\ref{secper:2}); we then describe the various
models suggested for their formation (Sect.~\ref{secper:3}) and the long term
evolution of BSSs in cluster environments (Sect.~\ref{secper:4}). Finally we compare
the expectations from the different models with the known observable
constraints and point out future theoretical and observational directions
to advance the field (Sect.~\ref{secper:5}) and summarise (Sect.~\ref{secper:6}). Some of
the subjects discussed in this review are explored in more details
in other chapters of this book; and we refer the reader to these chapters
when relevant. Though we discuss a wide variety of BSSs in different
environments, our main discussion will focus on BSSs in old OCs and
GCs which are best characterised; BSSs in other environments are discussed
more briefly.

\section{The Observed Properties of BSSs}
\label{secper:2}
Like any other stellar populations, BSSs are characterised through
a wide variety of properties. These could be divided between intrinsic
physical properties of the BSSs (mass, radii, composition, rotation,
variability, temperature, luminosity); physical and orbital characteristics
of multiple BSSs systems (binaries, triples); and the overall properties
of the BSSs population (frequency, multiplicity, radial distribution).
Another important division is between the directly observed BSS properties
--- e.g. colour-magnitude diagram (CMD) location --- vs. inferred properties
which require assumption dependent modeling (e.g. BSS mass). All of
these properties may differ in different environments where BSSs are
observed, and should therefore be discussed in the context of the
relevant environment. 

In the following we briefly discuss the observed properties of BSSs.
Cases where the relevant properties are not yet well characterised/understood
are specifically indicated in the table and text. An extended discussion
about the observed properties of BSSs can be found in Chap. 3 and 5.

\subsection{Physical Properties}
\label{secper:2.1}

\subsubsection{Masses }
\label{secper:2.1.1}
The masses\index{mass} of single BSSs are not known, and can only be inferred through
interpretation of their location in colour-magnitude diagrams as well
as spectroscopic data, in the context of stellar evolution\index{stellar evolution} models.
Detailed atmospheric models could potentially provide good constraints
on the mass, and such models provide mass estimates of up to twice
or more the turn-off\index{turn-off} mass for the brightest BSSs~\cite{sha+97}. Stellar
variability in SX Phe stars\index{SX Phe star} (all are BSSs in GCs) can also give various
clues on the matter, and provide mass estimates up to twice and even
three times the turn-off mass~\cite{nem+95}. 

However, given the complex
origin and stellar evolution of BSSs, and the inherent theoretical
uncertainties in these modeling such interpretation might not be very
reliable. In principle, the location of BSSs in the CMD shows them
to be hotter and more luminous than stars on the main sequence, leading to the
current interpretation of BSSs as stars more massive than the turn-off mass
of their environment. More reliable methods make use of the dynamics
of BSSs in multiple systems, where radial velocity measurements and/or
eclipses can provide additional information. Even those methods can
typically provide only partial data and/or constraints on the physical
properties of a specific BSS. In cases where a BSS mass was determined
dynamically (in double-lined spectroscopic binaries), it was found that it
was underestimated by 15\% compared with the mass inferred from stellar
evolution modeling of the CMD location (\cite{gel+12} and Chap. 3).

It is therefore premature to discuss a detailed mass function
of BSSs. In the following we therefore refer only to the range of BSS
mass inferred from the CMD, keeping in mind the potential large systematic
deviations of these masses from the real BSS masses. 

The CMD inferred mass function of BSSs in the OC NGC 188\index{NGC 188} \cite{gel+12} lies
in the range of 1.15--1.55~${\rm M}{}_{\odot}$ (see Chap. 3), i.e. $\Delta m=0.15-0.55$~${\rm M_{\odot}}$ more
massive than the cluster turn-off mass ($\sim$1~${\rm M_{\odot}}$); a
statistical estimate of the BSS masses based on orbital solution of
binary BSSs in the cluster suggest a comparable but slightly lower
mass range of 1.1--1.45~${\rm M}{}_{\odot}$~\cite{gel+12}. Among
field\index{field} BSSs, Carney,  Latham \& Laird \cite{car+05} find BSSs masses in the range 0.83--1.28~${\rm M_{\odot}}$,
i.e. $\Delta m=0.03-0.48$~${\rm M_{\odot}}$more than the turn-off mass
(0.8~${\rm M_{\odot}}$). In other words BSSs can be significantly
more massive than the turn-off mass, possibly requiring a large amount of
mass accumulated onto them from an external source. 

\subsubsection{Rotation}
\label{secper:2.1.2}
The rotation velocities measured for BSSs extend over a wide range,
showing both population of slow rotating\index{rotating star} and fast rotating stars (compared
with the background population; see \cite{lov+10,lov+13} and Chap. 5 and 3), with varying
distributions in different clusters. Systematic study of BSS rotational
velocities in different environments is still in its infancy, and
more data are needed before a clear interpretation of the data can
be done (e.g., rotational velocity dependence on cluster properties).
Relating these data to theoretical predictions is also premature,
given the contradicting theoretical results regarding BSS rotational
velocities (e.g., \cite{benz+87,leo+95}; see also Chap. 12) . More theoretical as well as observational
exploration is needed.

\subsubsection{Composition}
\label{secper:2.1.3}
Though BSS composition could provide important constraints on their
origin, e.g., showing pollution by accreted material from evolved stars, 
the available data is currently limited. We refer the reader to
Chap. 5 as well as an overview by \cite{lov+13}.
We will not discuss composition issues in this chapter.

\subsection{Population Characteristics}
\label{secper:2.2}

\subsubsection{Frequency}
\label{secper:2.2.1}
The overall frequency of BSSs in clusters is very small, but had typically
been measured in detail only in GC\index{galactic cluster} cores and in open clusters\index{open cluster}. Typically,
globular clusters contain a few, up to hundreds of BSSs, compared to
the large numbers of stars in these clusters (few $10^{5}-10^{6}$
stars), providing BSS fractions of the order of a few $10^{-5}-10^{-4}$.

Simulations of BSS formation in clusters (e.g., \cite{hyp+13}) suggest
that these fraction never become higher than these numbers even in the
early evolution of a GC. A few tens of BSSs have been found in old
open clusters such as M67\index{M67} and NGC 188\index{NGC 188}, providing a BSS fraction of
a few $10^{-3}$, i.e. much larger that that observed in GCs. Overall,
it appears that BSS frequency is inversely proportional to the stellar
density of the environment (for a more detailed discussion of these issues, see
Chap. 9).

\subsubsection{Multiplicity, Companion Type and Orbital Properties}
\label{secper:2.2.2}
Given the important role of binary or even triple companions in the
suggested models for BSS formation, BSS multiplicity\index{multiplicity} is one of their
most important properties. Unfortunately, it is difficult to characterise.
BSS multiplicity was studied in low mass field BSSs and in several open
clusters\index{open
cluster}, most notably M67\index{M67} and NGC 188\index{NGC 188} \cite{lat+07,gel+08}; much
less is known about the multiplicity of BSSs in GCs\index{globular cluster}.

\begin{figure}
\sidecaption
\includegraphics[width=75mm]{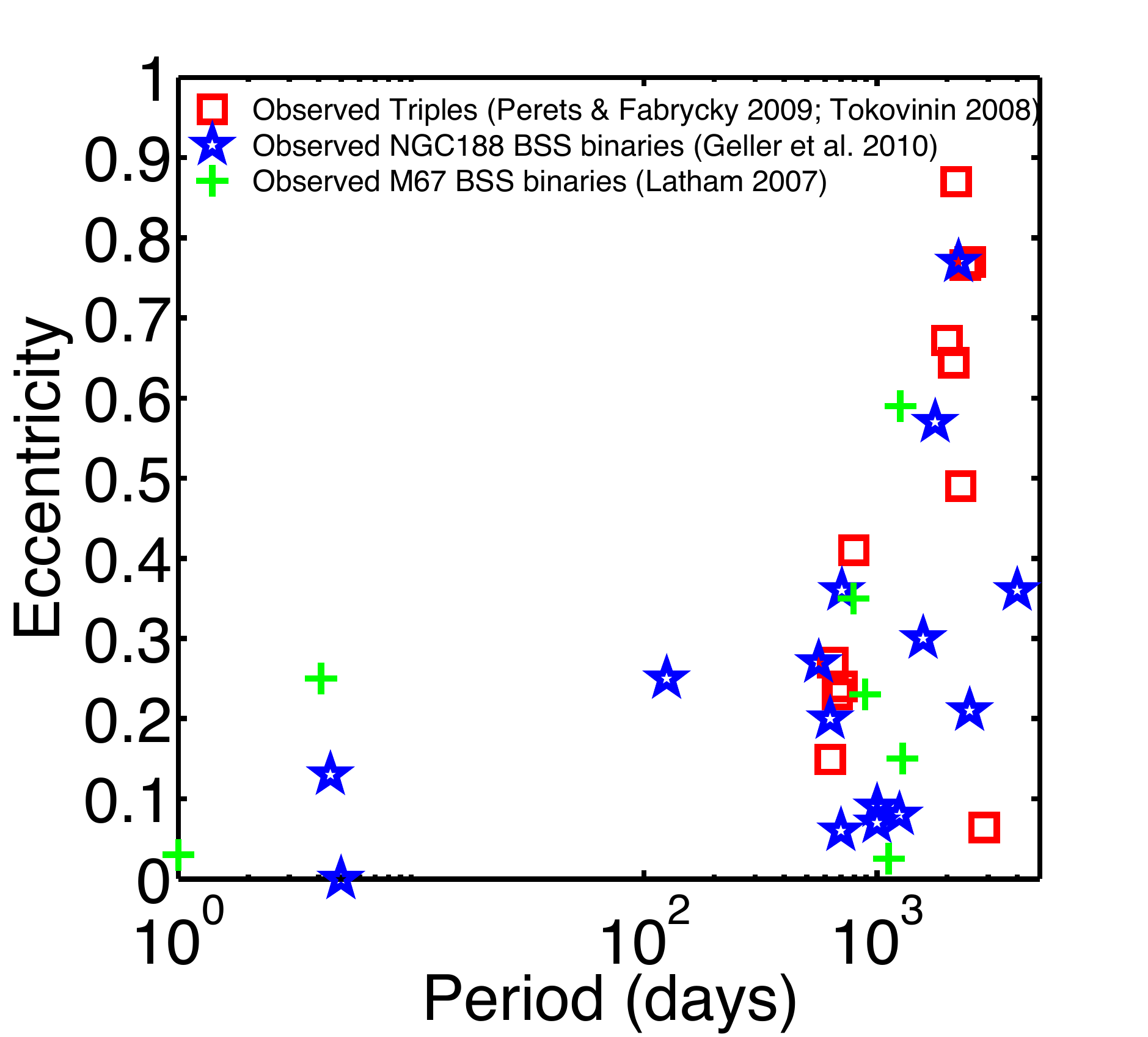}\caption{\label{fig:period-eccentricity}The period-eccentricity distribution
of BSS binaries in open clusters M67\index{M67} and NGC 188\index{NGC 188}, compared with the
outer orbits or field triples with short period inner binaries.}
\end{figure}

{\bf Field BSSs} Carney et al. \cite{car+05} used radial velocity measurements to study BSSs in the
field and find their binary fraction to be high --- consistent with
all of the observed BSSs having companions. Their analysis shows BSSs
binaries to typically have periods of 200--800 days, with low eccentricities
compared with field binaries at the same period range, but distinctively
not circular (see Fig. \ref{fig:period-eccentricity}). Statistical
analysis of the BSS companion masses show their mass to peak at $\sim0.6$~${\rm M_{\odot}}$, with none of the BSS binary companions directly
observed (all binaries were single-lined spectroscopic binaries\index{spectroscopic binary}). This was interpreted as
pointing to white dwarf\index{white dwarf} (WD) companions and therefore to a case C
mass transfer\index{Cases A, B, C
of mass transfer} scenario for the BSS formation. 

{\bf OC BSSs}\index{open cluster} The on-going effort to characterise the properties of BSS populations
in OCs, have provided us with very detailed knowledge\footnote{Currently, only for
two OCs but on going study will provide similar data for additional OCs
in the coming few years.}  about their
multiplicity and the orbital properties of BSS multiples (see Chap. 3 for a detailed discussion). The BSS populations in M67\index{M67} and NGC 188\index{NGC 188} show
many similarities. In both clusters radial-velocity studies show the BSS binary
fraction is much higher than the background stellar population
(60$\pm24$\% and 76$\pm22$\% for M67 and NGC 188, respectively), with the locus of
the period distribution extending between 700 days and 3000 days.
Though the upper limit is the observational limit, mostly due to the
long time baseline required, the high binary fraction of BSSs even
in this limited regime is much higher than that of field binaries
\cite{rag+10}, and points out to the important role of binaries.
Few binary BSSs are found at shorter periods, with both cluster showing
examples of peculiar double BSS binaries, and BSSs with more than twice
the turn-off mass of the OC, indicating the need for many body ($>3$) interaction
origin. The eccentricity distribution is distinctively not circular
and extending higher than that observed for field BSSs, but lower than
the background binary population in the clusters (\cite{gel+12};
see Fig.~\ref{fig:period-eccentricity}). Statistical analysis of
the companion mass shows it to peak at $0.55$~$M_{\odot}$ somewhat
similar to the case of field BSSs.

No systematic study of BSS binarity in GC\index{globular cluster} have ever been done. However,
variability surveys of several GCs suggest that the frequency of eclipsing
binaries among BSSs is much higher than that of field binaries \cite{mat+90,ruc+00}. 

%%%%%%%%%%%% FIGURE 1 %%%%%%%%%%%%%%
%%%%%%%%%%%% FIGURE 1 %%%%%%%%%%%%%%

\subsubsection{Radial Distribution}
\label{secper:2.2.3}
Some of the early studies of BSS radial distributions\index{radial distribution} in clusters have
already shown them to be centrally concentrated (in OCs, see \cite{mat+86}; in
GCs, see \cite{aur+90}). Later studies have revealed the existence
of a bi-modal radial distribution in some GCs and OCs (for the 
OC NGC 188, see \cite{gel+08}; in GCs, see \cite{fer+97} and Chap. 5), with an inner centrally concentrated
region followed by a dip in the distribution and a rise in the outer
parts. The existence of a bi-modal distribution appears to depend
on the cluster properties, and in particular the relaxation time, suggesting
a major role of mass segregation in determining the BSS radial distribution.
It is not clear whether other stellar populations show bi-modal radial
distribution, though observations of eclipsing binaries suggest that
short period binaries in the GCs $\omega$ Cen\index{$\omega$ Centauri} and 47 Tuc\index{47 Tucanae} may give
rise to such bi-modal distributions \cite{wel+04,wel+07,per+09}.

\subsubsection{Multiple Populations in GCs}
\label{secper:2.2.4}
Recently, it was shown that some GCs appear to host two distinct populations
of BSSs as observed in the CMD. The origin of such multiple populations
is yet to be studied in detail. We refer the reader to Chap. 5 for further discussion of this issue.

\section{Models for Blue Straggler Star Formation}
\label{secper:3}
All current models for BSS formation are based on the assumption that
these stars are rejuvenated through mass transfer\index{mass transfer}. The difference
between the models are the type of processes leading to such additional
mass accumulation. All models require an external source of material
and can be generally divided into three classes of mass transfer: merger\index{merger},
collision\index{collision} or accretion\index{accretion}. Mergers occur when two MS stars, typically
in a short period binary\index{binary star}, come into contact and eventually merge
together to form a more massive star containing most or all of the
the mass of the mergers binary. Collisions of two MS stars are a more
violent scheme where two MS stars form a merged star through a fast
dynamical encounter; these could occur through the collision of two
unrelated stars in a dense cluster\index{star cluster}, or possibly in unstable triple
stars\index{triple system} where two stellar companions collide. The more violent nature
of these events could produce different outcomes than the more gentle
merger processes, and can potentially produce BSS with different physical
properties. Finally, stellar companions could shed mass through winds\index{stellar wind}
or Roche lobe overflow\index{Roche lobe overflow} and the ensuing accretion 
can then rejuvenate the stars to become BSSs. 

There is a wide variety of stellar systems and different types of
evolutionary processes that could lead to these mass exchange scenarios;
in the following we discuss these models in more detail.

\subsection{Collisions in Dense Clusters}
\label{secper:3.1}
Early on, stellar collisions\index{collision} in dense clusters\index{star cluster} have been suggested
as a channel for BSSs formation. This channel could play a major role
in BSS formation in GCs\index{globular cluster} (e.g. \cite{cha+13,hyp+13}, and references therein)
and may contribute to the BSS population in the cores of OCs\index{open cluster} \cite{hur+05};
although it is not likely to serve as the dominant formation channel
in OCs \cite{leo96,hur+05,per+09,gel+13}. Obviously, this channel
is irrelevant for field BSSs, where physical collisions are extremely
rare. This formation channel is discussed in more details in Chap. 9; here we provide a brief discussion and focus
on the issue of binary BSSs in this context.

Collisional merger of stars is very efficient and conserves most of
the mass of both merged stars for low velocity encounters as expected
in OCs and GCs (\cite{benz+87}; large mass loss could occur at
impact velocities at infinity comparable to the escape velocity from
the stars), allowing them to form BSSs of up to twice the turn-off\index{turn-off} mass of
the cluster (or even more, if more than two stars collide). 

The rate of stellar collisions is strongly dependent on the number
density of stars in the cluster. Collisions can occur through the
direct physical collisions between single stars in the cluster, but
encounters between higher multiplicity systems are more likely to
mediate most physical collisions in dense environments \cite{leo89,fre+04,per11,lei+11,cha+13}.

It was therefore expected that a strong correlation between the collisional
parameter in GCs (see Chap. 9) and the specific frequency of
BSSs should exist. A correlation with the binary fraction, given their
role as collision mediators should also be apparent. However, though
observations do show a correlation with the GC binary fraction\index{binary fraction}, the
strongest correlation is found to be with the GC mass, while no correlation
is found with the calculated collisional parameter (see \cite{lei+13} and references
therein). Most interestingly, Chatterjee et al.~\cite{cha+13} have recently made detailed
simulations of the evolution of GCs\index{globular cluster} in their BSSs populations, and
found that binary mediated stellar collisions are the dominant channel
for BSSs formation in dense clusters. Moreover, they find a clear,
though weak correlation with the cluster collisional parameter\index{collisional parameter}. They
suggest that the calculated collisional parameter based on observational
analysis of GC properties is inaccurate, due to accumulated errors
in the various observational parameters; in fact, they find no correlation
between the intrinsic accurate collisional parameter calculated for
their simulated GCs and an ``observed'' collisional parameter obtained
by making use of a an observational-like analysis of the cluster properties.
These results may explain the conundrum in correlation between the
``observed'' GC collisional properties and the BSS population which
was debated over the last few years and can provide for various important
pointers for new observations. 

Binary-single and binary-binary\index{binary star} encounters\index{encounter} are very likely to leave
behind a binary BSS; Chatterjee et al.~\cite{cha+13} find that $\sim60$ \% of the
BSSs in their simulations are in binaries. Studies of binary-binary
and binary-single encounters \cite{leo+91,leo+92,dav95,fre+04} show
that binary-binary encounters leave behind BSS with long period binary
companions (most typical are at periods of $10^{3}-10^{4}$ days)
with an almost thermal eccentricity\index{eccentricity} distribution (average eccentricity
of $2/3$ and somewhat lower for the shorter period binaries). Detailed
hybrid Monte-Carlo models\index{Monte-Carlo model} coupled with few-body simulations\index{N-body simulation} of GCs
also account for the later evolution of the binaries, and show the
binary distribution peak at a few tenth to a few astronomical units; only a small
fraction ($<10$ \%) have small semi-major axis comparable with
typical eclipsing binaries\index{eclipsing binary} found in GCs. 

Finally, we note that none of the studies of GC evolution have accounted
for primordial triples\index{triple system} and their evolution. In addition, though dynamically
formed triples have been shown to form quite frequently in GCs, and
potentially play a non-negligible role in BSS formation \cite{2008I},
their long term evolution in GCs have not been studied. Given the
potentially important role of triples in mediating BSS formation \cite{2008I,per+09,lei+11},
this is an important direction for future theoretical studies of BSSs
and other exotica in GCs. 

\subsection{Binary Evolution}
\label{secper:3.2}
Binary stellar evolution (BSSE) was one of the first models suggested
for the origin of BSSs \cite{mcc+64}. This general term refers to
several possible scenarios and outcomes. In the BSSE model, the evolution
of a stellar binary leads to mass transfer from a star to its binary
companion, thereby increasing its mass, potentially a long time after
its formation. A stellar binary could merge in which case the final
product will include most of or all the mass of both companions;
alternatively the companion might shed mass through Roche-lobe overflow
(RLOF\index{Roche lobe overflow}; see Chap. 8), winds or wind-RLOF\index{wind Roche lobe overflow} (see Chap. 7)
thereby transferring part, or most of its envelope to the now rejuvenated
primary, leaving behind a white dwarf. Various aspects/sub-channels for the BSSE model
for the formation for BSSs are discussed in various other chapters
in this volume (Chap. 7, 8, 9, and 12); here
we provide a general overview for these the various sub-channels,
their differences and their implications in a more general context.
When applicable we will refer the reader to the relevant detailed
discussions in other chapters.

%\subsubsection{Case A/B/C/D (Wind Roch-lobe overflow) transfer}
%\label{secper:3.2.1}
Mass transfer\index{mass transfer} during BSSE is traditionally divided into three categories;
depending on the state of evolution of the mass donor interior \cite{kip+94}:
MT during the MS (case A\index{Cases A, B, C of mass transfer}), beyond the MS but before helium ignition\index{helium ignition}
(case B), or beyond helium ignition (case C). Eggleton \cite{egg06} refines
case A MT into many more sub-categories (not detailed here; we refer
the reader to \cite{egg06}), and redefines case B and C MT: Case
B is the situation where the mass donor is in the Hertzsprung gap\index{Hertzsprung gap},
with a mainly radiative atmosphere, and case C is the situation where
the mass donor is on the giant branch\index{red giant}, and therefore have a mainly
convective envelope\index{convective envelope}. We will not discuss these scenarios in depth,
but rather remark on their main implications for BSS and its companion,
and focus only on BSS formation during MT. 

%%%%%%%%%%%% FIGURE 2 %%%%%%%%%%%%%%
\begin{figure}
\includegraphics[width=119mm]{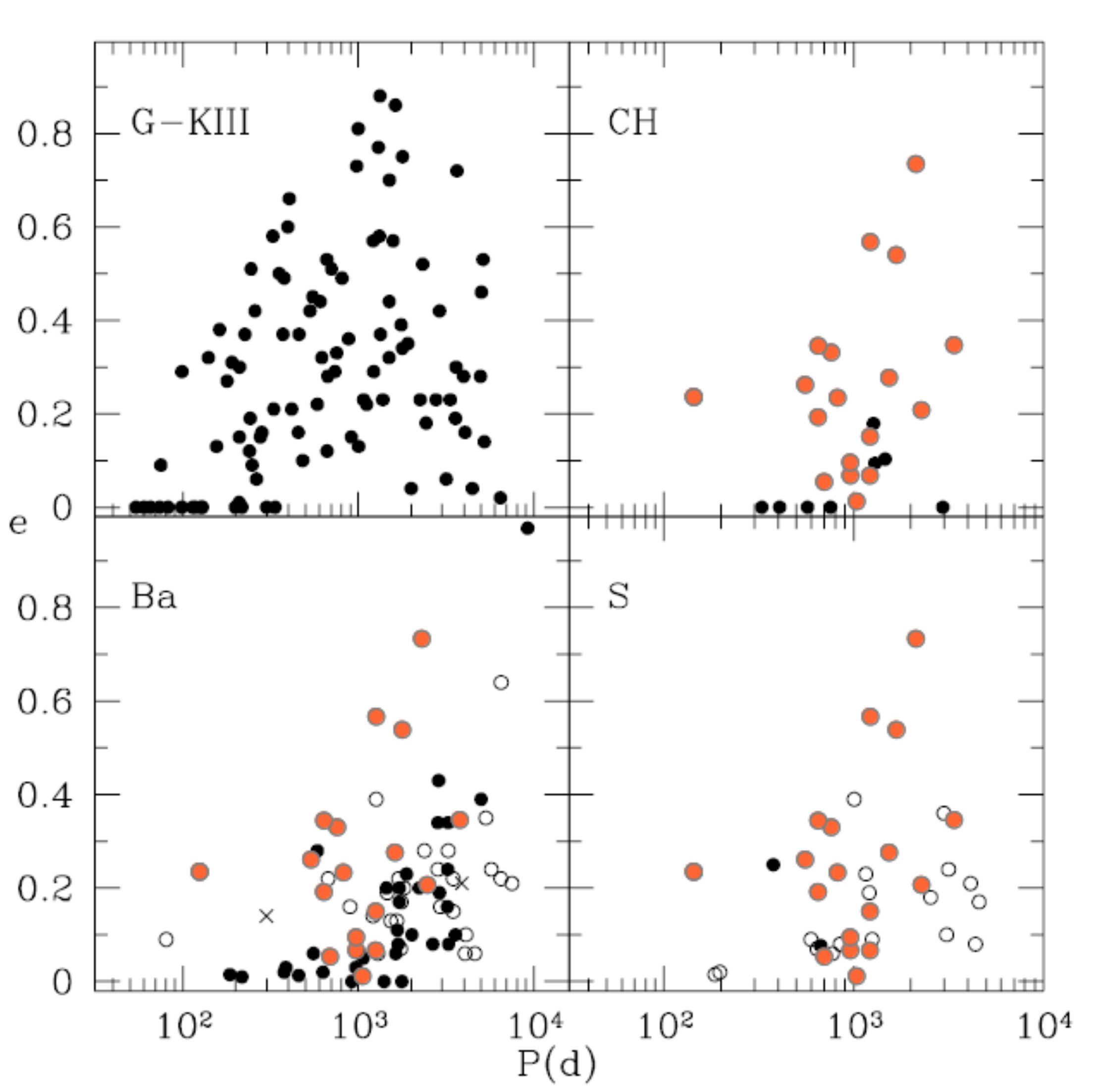}\caption{\label{fig:period-eccentricity-all}The period-eccentricity\index{orbital period} distribution
of several populations of evolved binaries, compared with the BSS binaries
in M67\index{M67} and NGC 188\index{NGC 188} (adapted from \cite{jor+98}). Upper left panel:
Binaries involving G and K giants in open clusters \cite{mer+96};
Upper right panel: CH stars\index{CH star} \cite{mcc+90}; Lower right panel: S
stars\index{S
star} \cite{jor+98}. BSS binaries in the open clusters M67\index{M67} and NGC
188\index{NGC 188} are shown as large (orange) filled circles. Note that although
BSSs were likely to accrete\index{accretion} more mass that the various type of polluted
stars, their eccentricities\index{eccentricity} are much higher than those of the polluted
star binaries, even at short period of a few hundred days (the most
comparable are the Ba stars\index{barium star}; however, these are stars far more massive
than the BSS binaries, whereas the CH stars with more comparable masses
show significantly lower eccentricities. In other words, it appears
than mass transfer does induce circularisation\index{tidal circularisation} of binaries (though less efficiently
that the typical theories suggest), as evident from the lower eccentricities
observed for the polluted stars. The origin of the higher eccentricities
at low periods for the BSS binaries (which accrete more mass than polluted
stars) is therefore inconsistent with the overall period-eccentricity
distribution of OC BSSs, suggesting that at most only small fraction
of them can be explained by case C mass transfer\index{mass transfer}. Mass transfer may still explain the origin
of the lowest eccentricity BSS binaries observed; whereas the rest
of the BSS binaries might be explained by the triple origin; see Fig.~\ref{fig:period-eccentricity} for comparison of period-eccentricity distribution in triples.}
\end{figure}
%%%%%%%%%%%% FIGURE 2 %%%%%%%%%%%%%%

MT scenario are generally thought to lead to orbital circularisation\index{circularisation}.
However, binary systems thought to be produced through MT many time
show distinctively larger eccentricities\index{eccentricity} (see Fig. \ref{fig:period-eccentricity-all}),
suggesting that our understanding of this process is incomplete. Some
studies suggests scenarios where higher eccentricities are kept (see \cite{der+13} and references therein).

{\bf Case A MT}: In order for a MT to occur during the MS, the
initial binary separation must be small enough for the RLOF\index{Roche lobe overflow} to ensue
(at most a few solar radii), given the compact radius of the companion
MS star. Various processes could lead to such a a close configuration,
involving magnetic braking\index{magnetic braking}, tidal evolution\index{tidal evolution} or possibly also affected
by perturbations from a third companion in a triple\index{triple system} (see Sec.~\ref{secper:3.3}
for the latter). In dense stellar clusters\index{star cluster}, interactions with other
stars in the cluster may also results in shortening the binary period,
exciting the binary eccentricity\index{eccentricity} (thereby leading to a smaller peri-centre
approach where effects from the the first processes can become significant),
or even exchange of companions to produce a new shorter period binary. 

RLOF\index{Roche lobe overflow} on the MS will increase the gainer mass, making it a BSS. Such
evolution could lead to evolution into contact\index{contact binary} configuration (during
the evolution they system could get in and out of contact, depending
on the specific case), and possibly merge. A merged system will form
a massive BSS, an unmerged system would form a BSS with a short period
companion, which could be long lived (1--2 Gyr); contact configuration
might be observed as W UMa\index{W UMa star} type eclipsing binaries. Case A MT therefore
lead to either single massive (on average, given the complete MT)
BSS, or to a less massive (on average, given the only partial MT) BSS
with a short period hydrogen burning companion (note however that
the companion could be affected by the interaction, its appearance
not necessarily resembling a MS star). 

{\bf Case B MT}: In this case the evolution into RLOF\index{Roche lobe overflow} occurs only
following the companion evolution to Hertzsprung gap\index{Hertzsprung gap}, in longer period
binaries (few to tens solar radii). For significant accretion to occur
leading to BSS formation, the system should not go through a common
envelope\index{common
envelope} stage which will eject the envelope rather than lead to mass
growth. Therefore, scenarios leading to a BSS formation will leave
behind a BSS with an intermediate period (few$\times1-10$ days) and
a helium WD\index{helium white dwarf} companion, following the mass donor stripping; see \cite{lan+97}
for a detailed example of such a scenario used to explain the origin
of the 1040S system in the OC M67\index{M67}. In principle the BSS binary could
also be observed during the accretion phase, but only for a relatively
short time of a few hundred Myr. The BSS could be quite massive, though
0.1--0.4~${\rm M_{\odot}}$ of the final system mass will reside in
the helium WD. Such BSSs are not likely to be observed as eclipsing
binaries\index{eclipsing
binary} (due to the small radius of the WD companion, as well as
the expected wide separation), unless observed during the accretion
phase.

{\bf Case C MT}: In this case the donor star is already quite
evolved, with a large radius. The initial binary period is therefore
expected to be in the range of tens to a few hundreds or even 1--2
thousand days. In systems with up to a few hundred days period, a
BSS could be formed through accretion\index{accretion}, with up to $\Delta{\rm M}\sim0.2\,{\rm M_{\odot}}$
above the turn-off\index{turn-off} mass. Stellar evolution calculations (e.g. \cite{che+08})
suggest that binaries with larger periods can only produce BSSs very
close to the turn-off mass, which are not considered as BSSs in current observational
criteria. After its evolution the donor star will become a CO WD\index{CO white dwarf}\index{white dwarf},
with typical mass of $0.6$~${\rm M_{\odot}}$. 

{\bf Case D MT (wind RLOF\index{Roche lobe overflow})}: Binaries with wide separations (typically
$>4$ AU) will not evolve through RLOF, as the primary star do not
fill its Roche-lobe. MT could still occur through the accretion of
slow wind\index{stellar wind} material ejected from an asymptotic giant branch\index{asymptotic giant branch star} (AGB) star. Simple calculations using Bondi-Hoyle accretion\index{Bondi-Hoyle accretion} model suggest such
accretion is inefficient at large separations ($<15$ \%). However,
recent hydrodynamical simulations\index{hydrodynamical simulation} showed that at binary periods of
$\sim2,000-10,000$ days, the wind can be focused by the accreting
star and the accretion\index{accretion} efficiency can be as high as $45$\% \cite{aba+13}.
It is therefore potentially possible to form even massive BSS ($\Delta M>0.4$)
through this process. The leftover companions should be CO WDs\index{CO white dwarf}\index{white dwarf}, following
the regular evolution of the donor star after the AGB. 

\subsection{Triple Evolution}\index{triple system}
\label{secper:3.3}
%\label{sub:triple}

Binary stellar evolution has been suggested early on as a channel
for BSS formation through MT and mergers. In recent years, however,
it was realised that triple stars can have an important, and sometime
major role in affecting binary evolution. Interestingly, the evolution
of triple system could mediate the production of BSSs through various
different channels thereby allowing for the formation of BSSs through
mergers, MT and even collisions. In the following we discuss these
various channels, focusing mostly on secular evolution through Kozai-Lidov
cycles\index{Kozai 
cycle}, coupled with tidal friction\index{tidal friction} (KCTF). 

\subsubsection{Secular Evolution Coupled with Tidal Friction}
\label{secper:3.3.1}
%\label{sub:secular}

As discussed above, one of the channels for BSS formation is case A
MT, through which stars in short period binary systems can transfer
mass or even merge on the MS, thereby leading to the formation of
massive BSSs. Such short period binaries can not easily form following
the evolution of pre-MS binary, since the pre-MS radius of stars can
sometime be larger that the short period binary\index{binary star} separation. It was
therefore suggested that the formation of short period binaries is
mediated by triple dynamical evolution \cite{1998KEM,2001EK,2006EK,2007FT}
coupled with tidal friction processes, as we shall discuss in the
following. Following this, Perets \& Fabrycky~\cite{per+09} suggested that triple stars
could serve as natural progenitors for BSSs, and in particular could
explain the existence of eccentric and wide orbit BSS binaries observed
in OCs, as triples in which the inner binary have merged, leave behind
a BSS with a long period companion. 

Stable triple systems require a hierarchical\index{hierarchical triple system} configuration in which
two stars orbit each other in a tight ``inner binary'', and the
third star and the inner binary orbit their common centre of mass
as a wider ``outer binary''. Such triples are long lived, but secular
evolution can sage their orbital inclination and eccentricity. A particularly
important change was discovered by Kozai~\cite{1962K} and Lidov~\cite{1962L}
in the context of solar system triples (Sun-asteroid-Jupiter or Sun-Earth-satellite).
They found that if the inner binary initial inclination relative to
the outer binary orbit is high enough, secular torques will cause
its eccentricity\index{eccentricity} and inclination\index{inclination} to fluctuate out of phase with one
another, leading to periodic high amplitude oscillations in the inclination
and eccentricities of the triple inner binary; these are typically
termed ``Kozai oscillations''\index{Kozai cycle}. Lidov~\cite{1962L} noted that the large
oscillations in the amplitude of the inner binary eccentricity might
even lead to collision\index{collision} between the inner binary members. Collisions
were prominent also in the first application of these dynamical concepts
to triple stars. Harrington~\cite{1968H} noted that large initial inclination
($i_{c}\lesssim i\lesssim180^{\circ}-i_{c}$, for a ``Kozai critical
angle'' of $i_{c}\approx40^{\circ}$), leads to large eccentricities,
which could cause a tidal interaction, mass loss, or even collision
of the members of the inner binary. Thus, Harrington~\cite{1968H}  reasoned that
a triple star system with an inner binary mutually perpendicular to
the outer binary should not exist for many secular timescales. However,
as noted by \cite{1979MS}, the inner binary stars, while coming close to
one another, will not merge immediately; instead, the tidal dissipation\index{tidal dissipation}
between them shortens the semi-major axis\index{semi-major axis} of the inner binary during
these eccentricity cycles. They suggested that such inner binaries
could therefore attain a very close configuration, in which mass transfer
and accretion could occur, possibly forming cataclysmic variables\index{cataclysmic variable}
or binary X-ray\index{low-mass X-ray binary} sources. Kiseleva, Eggleton \& Mikkola~\cite{1998KEM} were the first to show that
coupling of tidal friction\index{tidal friction} to the high amplitude Kozai-Lidov secular
evolution could be a highly efficient mechanism for the formation
of short period binaries, as was later studied by \cite{2006EK,2007FT};
we refer to this mechanism as Kozai cycles and tidal friction (KCTF;
note that the last stages of the merger\index{merger} are likely to be induced by
magnetic braking\index{magnetic braking}). 

Observations of short period binaries showed that more than 90\% of
short period F/G/K binaries with periods of $P<3$ days --- consistent
with 100\%, when considering completeness --- have a third companion.
The fraction gradually decreases to $\sim30$ \% at $6\le P\le30$
days, consistent with the overall background level ($\sim30$ \% of
all F/G/K binaries are triples \cite{rag+10} ). These observations
suggest that all short period binaries $(P<3-6$ days) form in
triple systems, giving credence to the KCTF formation scenario. 

\begin{figure}
\includegraphics[width=119mm]{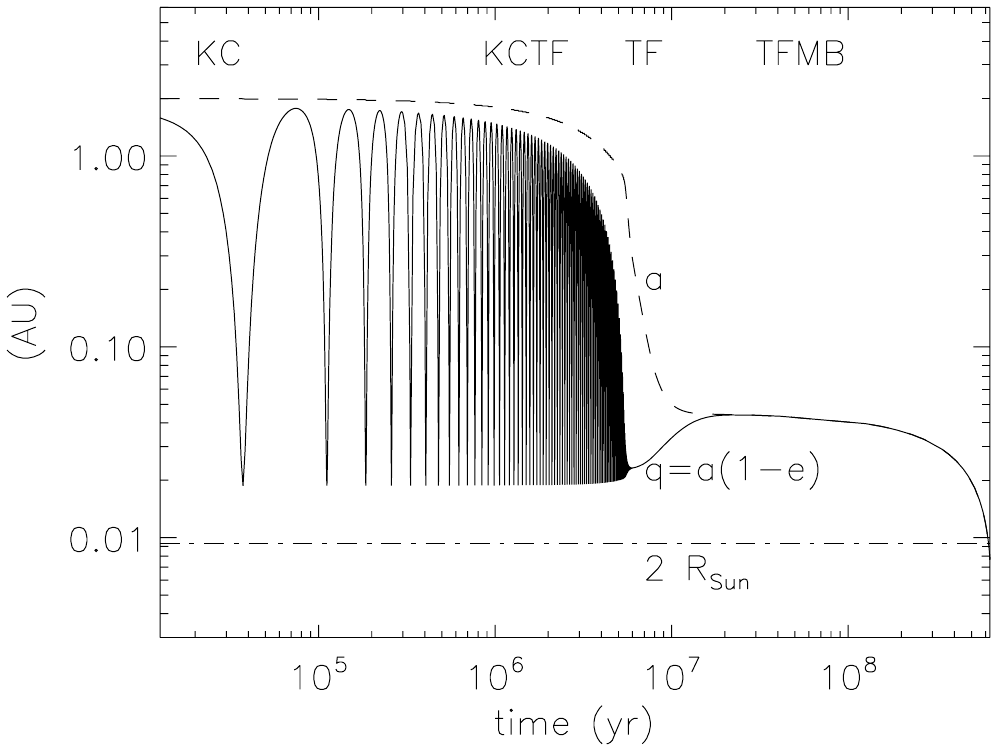}\caption{\label{fig:KCTF}Merger of the two stars of an inner binary, accomplished
by a combination of Kozai cycles, tidal friction, and magnetic braking.
(Reproduced from \cite{per+09} by permission of the AAS)}
\end{figure}

Perets \& Fabrycky~\cite{per+09} have taken the next logical step; if short period
binaries are formed through KCTF evolution in triples, and short period
binaries are typical progenitors of BSSs, then triples could be the
natural progenitors of BSSs. In Fig.~\ref{fig:KCTF} we show an example
for the KCTF scenario and the formation of short period binaries that
wild later merge. 

%%%%%%%%%%%% FIGURE 3 %%%%%%%%%%%%%%
%%%%%%%%%%%% FIGURE 3 %%%%%%%%%%%%%%

Most interesting, this basic scenario provides a wealth of unique
observable predictions. These could be summarised by the following: 
\begin{enumerate}
\item Merged short period binaries in triples will leave behind a BSS and
a binary (originally the third) companion. Therefore the binary
frequency of BSSs should be high, with typically wide periods, otherwise
the original triple would not have been stable. 
\item BSSs in short period binaries should still have a third companion on
a wider orbit.
\item The BSS mass (and luminosity) can be high, and as much as twice the
turn-off\index{turn-off} mass in the case a full merger\index{merger} of the inner binary; this is to
be compared with BSSs in wide orbit binaries from case B/C MT, where
a significant mass is left in the WD companion, and the MT efficiency
is not expected to be high (typically at most $0.2$~${\rm M_{\odot}}$
can be transferred to the mass accretor).
\item The period-eccentricity\index{orbital period}\index{eccentricity}\index{period-eccentricity diagram} diagram of BSS binaries should show strong
similarities to the period-eccentricity diagram of the \textbf{outer}
binaries of triples with inner short period binaries. Such a comparison
is shown in Fig. \ref{fig:period-eccentricity}; it appears that such
binaries should have long periods ($\gtrsim500$ days) and eccentricity
distribution comparable to that of field binaries with similar periods.
\item The wide orbit companion of BSSs could be a MS star in contrast with
case C or case B MT which might produce a wide orbit BSS binary, but
the companion in that case must be a CO (typical mass of 0.5--0.6~$M_{\odot}$; case C) or a helium white dwarf\index{CO white dwarf}\index{white dwarf}\index{helium white dwarf} (of $015-0.45$~${\rm M_{\odot}}$;
case B). 
\end{enumerate}

Recent studies have shown that inclined\index{inclination} triple systems with smaller
hierarchies (i.e. the outer to inner period ratio is small, of the
order of a few times the critical ratio for the system stability),
show quasi-periodic oscillations\index{oscillation}, similar to Kozai cycles, but
more chaotic, and which can lead to large eccentricity change on dynamical
timescales\index{dynamical
timescale}, and to higher eccentricities\index{eccentricity} (\cite{ant+12,kat+13},
see also \cite{ham+13} for a related study). Such behaviour extends
the Kozai-induced mergers to the small hierarchy regime, and could lead
to direct collisions even without significant tidal evolution\index{tidal evolution}, i.e.
triple dynamical evolution could lead to physical collisions\index{collision} and not
only mergers/case A MT.

\subsubsection{Collisions in Destabilised Triples }
\label{secper:3.3.2}
Recently, \cite{per+12} have suggested that stellar evolution and
mass loss from the inner binaries of triples can lead to the triple
destabilisation and the occurrence of collisions and close tidal encounters
between the triple components. Such triple evolution could therefore
lead to the formation of BSS when two MS stars collide. However, it
is not clear that such evolution could produce a significant population
of BSSs, and more detailed triple stellar evolution studies are needed
(see also \cite{ham+13} for a population synthesis calculations,
but limited to wide binaries of $a>12$ AU; much beyond observed binary
BSS separations in OCs). 

\subsubsection{Accretion onto a Binary from a Third Companion}
\label{secper:3.3.3}
Another interesting evolutionary scenario is the case where a third
companion in a triple evolves and sheds its mass on the inner binary.
Such a scenario have not yet been studied in details. In principle,
the transferred mass might form a common envelope\index{common envelope} around the inner
binary, leading to its in-spiral and the formation of a short period
binary that could later evolve to form a BSS through a case A MT. In
addition, or alternatively, the mass can accrete onto the inner binary
components potentially rejuvenating both of them and forming a double
BSS binary with a WD third companion on a wide orbit. Whether these
scenarios are physically plausible is yet to be confirmed in detailed
studies (see also Chap. 4).

\section{Long Term Dynamical Evolution of BSSs in Clusters}\index{star cluster}
\label{secper:4}

\subsection{Mass Segregation in Clusters}
\label{secper:4.1}
BSSs are more luminous and bluer than the background population in
which they are detected, suggesting their mass to be higher than the
turn-off\index{turn-off} mass. In particular, this would make BSSs the most massive stars
in stellar clusters, beside neutron stars and black holes. Dynamical
friction (see Chap. 10 and 9) leads to the segregation\index{mass segregation}
of the more massive stars into more centrally concentrated distribution
compared with the background population of lower mass stars. Currently
observed evolved stars (red giants\index{red giant} and horizontal branch\index{horizontal branch star} (HB) stars)
have evolved off the MS only relatively recently, i.e. their mass
during most of their evolution was close to the currently observed
turn-off mass of the cluster, and they were among the long lived most massive
stars in the cluster. Since these stars are also more massive than
the typical stars in the cluster they should also be mass segregated,
though slightly less than BSSs. 

BSS radial distribution\index{radial distribution} in clusters is typically compared with that
of HB or red giant stars. However, given the relatively small difference
in mass between this background population and the BSSs, it might be
surprising that BSSs appear to be more segregated than these populations.
In particular, Mapelli et al. \cite{map+04} and, more recently, Ferraro et al. \cite{fer+12} have
suggested that the bimodal radial distribution in clusters is due
to the segregation of the BSSs compared with the HB stellar population.
They provided a simplified formula for the position in the cluster
at which the timescale for BSS mass segregation is comparable to the
cluster lifetime, and showed that it appears to be consistent with trough
in the bimodal distribution of BSSs, showing this as evidence for
this process. However, if one were to use the same approach on red
giant\index{red giant} stars, taking their mass to be slightly smaller than the turn-off
mass of the cluster (as expected for most of their evolution), one
finds that they should similarly show a trough not far from the location
of the expected BSS trough. In other words a relative comparison of
the two populations should not have shown a significant trough in
the relative populations just due to mass segregation. The simple
theoretical interpretation therefore appears to be discrepant with
the observations. This might arise from the various assumptions made.
For example, the assumed mass of the BSSs was taken to be 1.2~${\rm M_{\odot}}$,
however, BSSs might have a much higher binary fraction than that of
of the background population, consistent with several of the BSS formation
channels. BSSs are therefore likely to have a binary companion, and
the mass of the BSS system should therefore typically be the mass of
a binary rather than the assumed mass of a single BSS. This might remedy
the discrepancy described above, and not less important point out
that the observed GC BSSs are likely to be binaries; an important clue
about their origin. Other problem may arise from not accounting for
the different lifetimes of the BSS population and that of the compared
population (M. Giersz, priv. comm. 2012).

\subsection{Dynamical Evolution of BSSs Binaries}
\label{secper:4.2}
BSSs can form through one of the evolutionary channels discussed above,
leaving them either as single stars or in a multiple system, which
could later change. In clusters their initial configuration could
dynamically evolve through encounters with other stars in the clusters.
Hurley et al.~\cite{hur+05} provide a detailed analysis of such later dynamical
evolution observed in simulations of an open cluster\index{open cluster}, and present
the diverse possibilities and outcomes of such evolution. The complicated
dynamical evolution of binaries in cluster is beyond the scope of
this review, but we will briefly discuss some of the main results
relating to BSSs. Encounter between multiple systems typically lead
to the more massive stars residing in the binary and the least massive
stars being ejected. BSSs are the most massive stars in their host
cluster and therefore have a higher probability to be exchanged into
binaries, even if they originally formed as single BSSs. Detailed analysis
of the evolution of open and globular clusters\index{globular cluster} showed that the BSS
binary fraction in these simulations were high, mostly due to their
formation channels, but their later dynamical evolution kept them
in binaries, and the majority of BSSs are found in binaries at the
end of the simulations \cite{hur+05,cha+13,hyp+13}.

Dynamical encounters\index{encounter} between binary and single stars typically leave
behind relatively eccentric binaries. Hurley et al.~\cite{hur+05} find the period-eccentricity
diagram\index{period-eccentricity diagram} of BSS binaries in their simulation to be inconsistent
with observations. In particular, most of the wide binaries of 200--700
day period were typically formed though dynamical encounters,
leaving behind highly eccentric binaries, whereas most observed BSS
binaries reside in larger periods ($>500$ days) and much less eccentric
orbits (see \cite{per+09} and \cite{gel+12}).

\section{Blue Straggler Stars: Observations vs. Theory}
\label{secper:5}
The various theoretical expectations and observational results discussed
above are summarised in table \ref{tab:BSS-obs} and \ref{tab:BSS-thoery}.
In table \ref{tab:BSS-obs} we show a summary of the observed properties
of BSSs in different environments, to be compared with the summarised
theoretical predictions shown in table \ref{tab:BSS-thoery}. The latter
include single processes, as well as results of large simulations
of GCs \cite{cha+13,hyp+13} and OCs \cite{hur+05,gel+13} which
include both dynamics and stellar evolution\index{stellar evolution}, but do not include triple
stars.

In the following we briefly discuss the theoretical expectations vis
a vis observations in the different environments.

\subsection{Globular Clusters}\index{globular cluster}
\label{secper:5.1}
Recently extensive Monte-Carlo simulations\index{Monte-Carlo simulation} of GC evolution \cite{cha+13,hyp+13}
provided for the first time detailed predictions regarding the population
of BSSs in these environments. These simulations include simplified
stellar evolution prescriptions and detailed account of binary dynamics
and interactions. These simulations do not account for primordial
or dynamically formed triples\index{triple system}. The detailed simulations provide the
overall evolution of the BSS population in GCs, but here we will focus
on the final outcome at the typical age of observed GCs ($\sim$12
Gyr). The simulations suggest that most of the currently observed
BSSs today result from direct physical collisions\index{collision} during binary-single
and binary-binary encounters\index{encounter}. The total number of BSSs in the simulations
is consistent with the observed numbers of BSSs in GCs. The BSSs are
centrally concentrated, but can show a bi-modal radial distribution\index{radial distribution}.
The majority ($\sim60$ \%) are in binaries with a wide orbital distribution
between a few days and a few thousand days, but mostly distributed
at $\sim100-1000$ days, with a small fraction ($\sim$10 \%) at short
periods. We conclude that these results are generally consistent with
the observed known properties of GC BSSs\index{globular cluster}, and can explain the correlation
between the binary fraction\index{binary fraction} and the BSS fraction (mostly due to the
dominance of collisions in binary encounters). Unfortunately, detailed
knowledge of the BSS properties such as exist for a few open clusters\index{open cluster}
precludes more detailed comparison (e.g., with BSS binary frequency
and binary orbital properties). It is also not yet clear whether these
models can explain the observed correlation of the normalised BSS fraction
with the mass of the GCs \cite{lei+13}. In addition, the existence
of a significant fraction of observed eclipsing\index{eclipsing binary} BSSs, as discussed
above, suggest a large fraction of the BSSs are in short period binaries\index{binary star}
with MS companions, inconsistent with the theoretical models. Such
short period BSS binaries may arise in the triple evolution scenario
where the induced formation of a short period inner binary leads to
case A MT, and the BSSs to be observed as eclipsing binaries. However,
at this point the data of eclipsing binaries is very sparse and detailed
dedicated observational study of GC eclipsing binaries as well as
theoretical study on the role of triples in GCs is required in order
to resolve this issue.

\subsection{Open Clusters }\index{open cluster}
\label{secper:5.2}
Detailed hybrid $N$-body/stellar evolution\index{N-body model}\index{stellar evolution} simulations by \cite{hur+05}
and \cite{gel+12} were done for old OCs similar to M67\index{M67} and NGC 188\index{NGC 188}
for which detailed data observations exist. A detailed overview of
the observations and simulation results can be found in Chap. 3. Overall they find that the
number of BSSs in the simulation is less than a third of that observed
and they can not reproduce the bimodal radial distribution of BSSs
observed in NGC 188; moreover the BSSs in the simulations are much
more centrally concentrated than the observed ones. The BSS binary
fraction in the simulations is only $\sim15$ \%, but a fifth of of
the observed value. To quote these authors 
\begin{quotation}
the deficiency
in number of BSSs, the low frequency of detectable binaries among those
that are formed, and the lack of a bimodal BSS radial distribution\index{radial distribution}
are striking failures of the model compared with the observations.
\end{quotation}

Nevertheless, Mathieu and Geller suggest that at least the orbital
properties of the BSS binaries in the simulations are consistent with
the observed BSS binary properties, reproducing the long periods of
the BSSs, non-circular orbits and companion mass distribution. Since
most of these binaries arise from case C MT scenario, they conclude
that this is the likely main mechanism for the BSS production in OCs,
irrespective of the other failures that might be remedied with better
stellar evolution models. However, case C MT that lead to the long
orbital periods with high eccentricity\index{eccentricity} is highly inefficient in transferring
mass, producing BSSs with typical $\Delta{\rm M}=\sim0.1$~${\rm M}{}_{\odot}$,
smaller than typically inferred from the CMD location of the BSSs.
We therefore conclude that case C MT may explain a fraction of the
observed BSSs, but it is difficult to see how it can explain the majority
of the OC BSSs. That being said, our current understanding of MT in
binaries is very limited, and future studies may show that more efficient
MT can occur (e.g., some form of case D MT). Evolution of primordial
triples may help produce the observed binary BSSs, not only naturally
explaining the long orbital periods and high eccentricities (see Fig.
\ref{fig:period-eccentricity}), but also explaining the very high
binary fraction, and the high mass of the BSS inferred from the CMD.
Detailed cluster simulations which include significant fraction of
triples and detailed accounting for KCTF are needed in order to check
whether a significant number of BSSs can indeed form in this way. 

\begin{figure}
%\sidecaption
\begin{center}
\includegraphics[width=58mm]{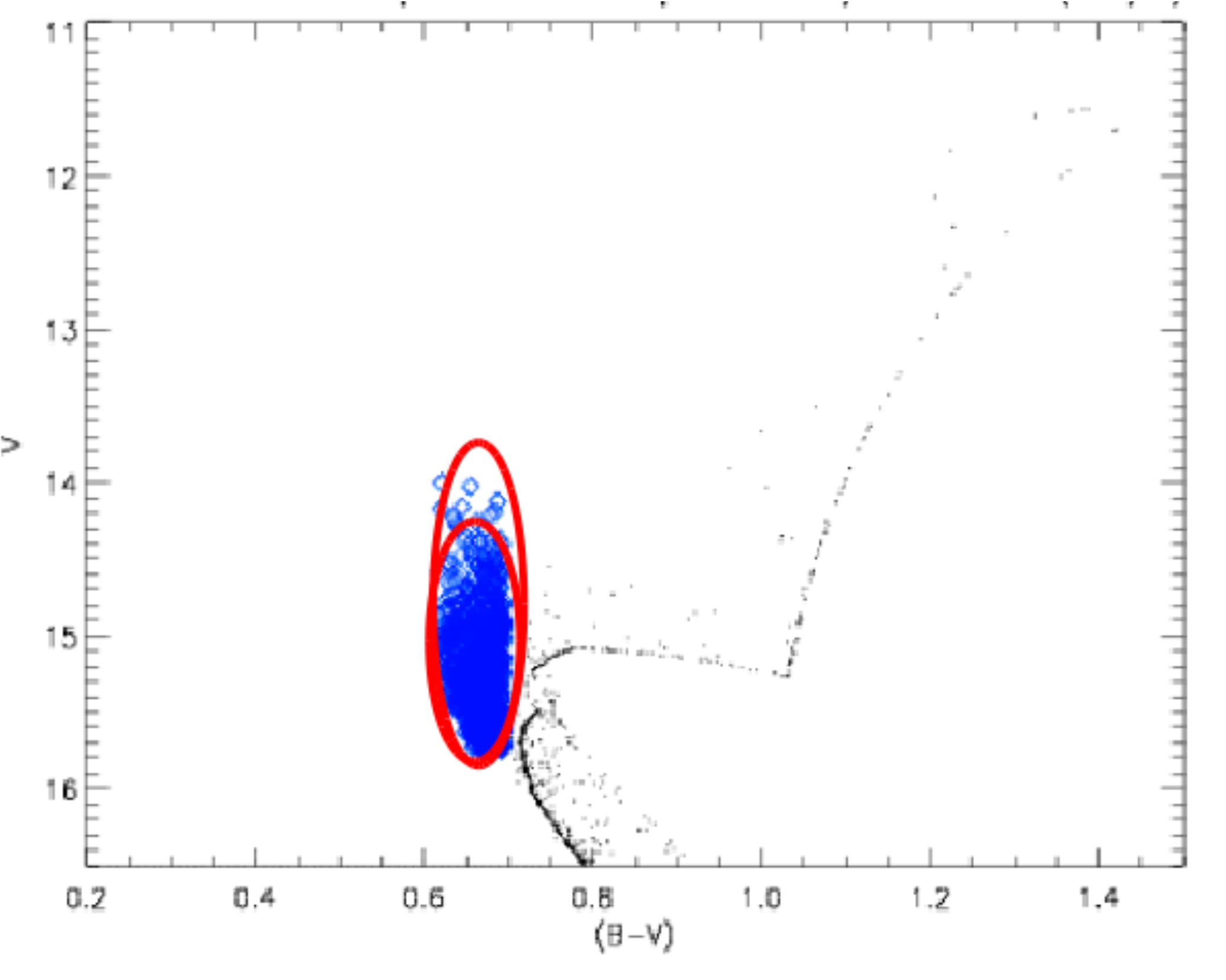}
\includegraphics[width=58mm]{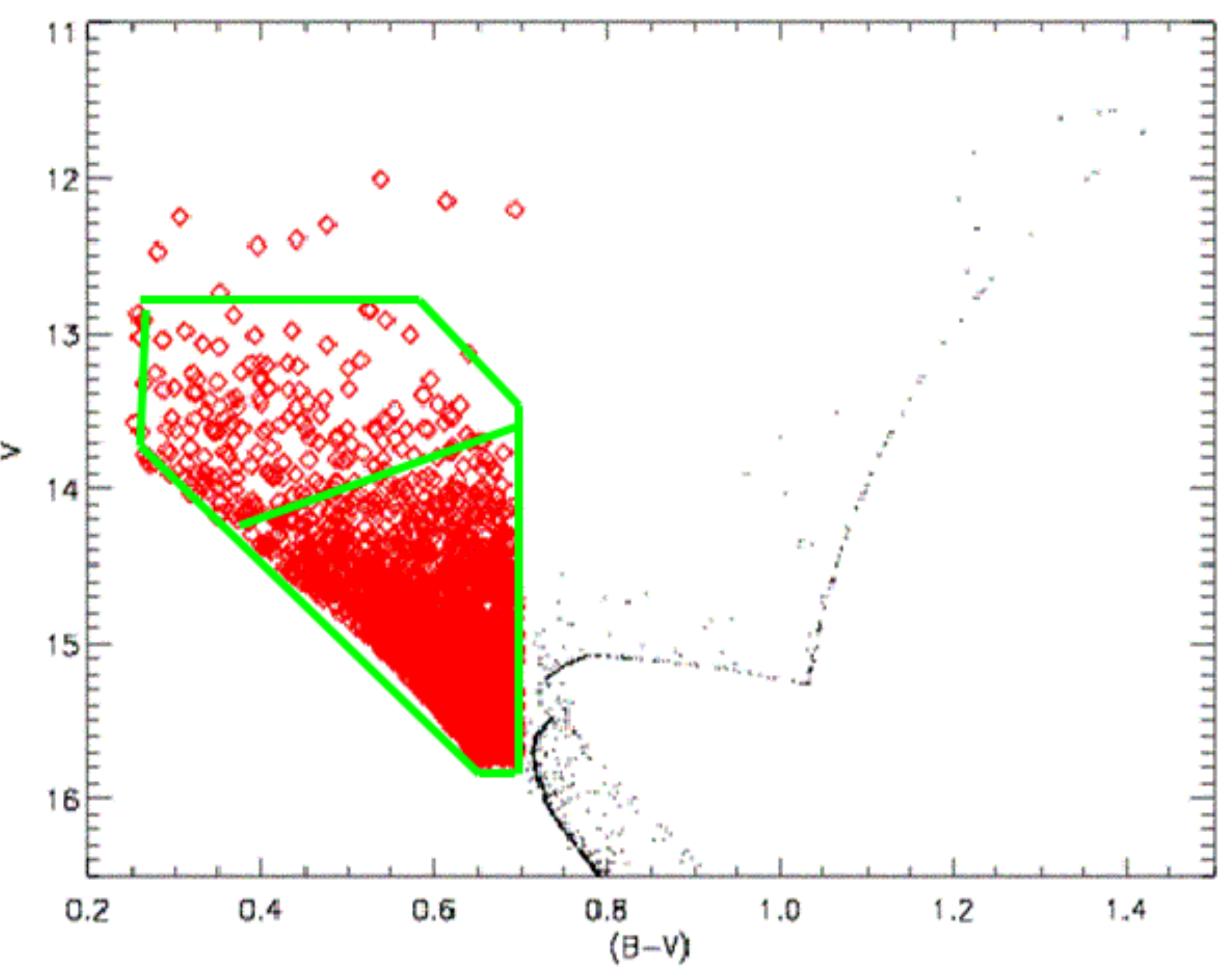}
\includegraphics[width=58mm]{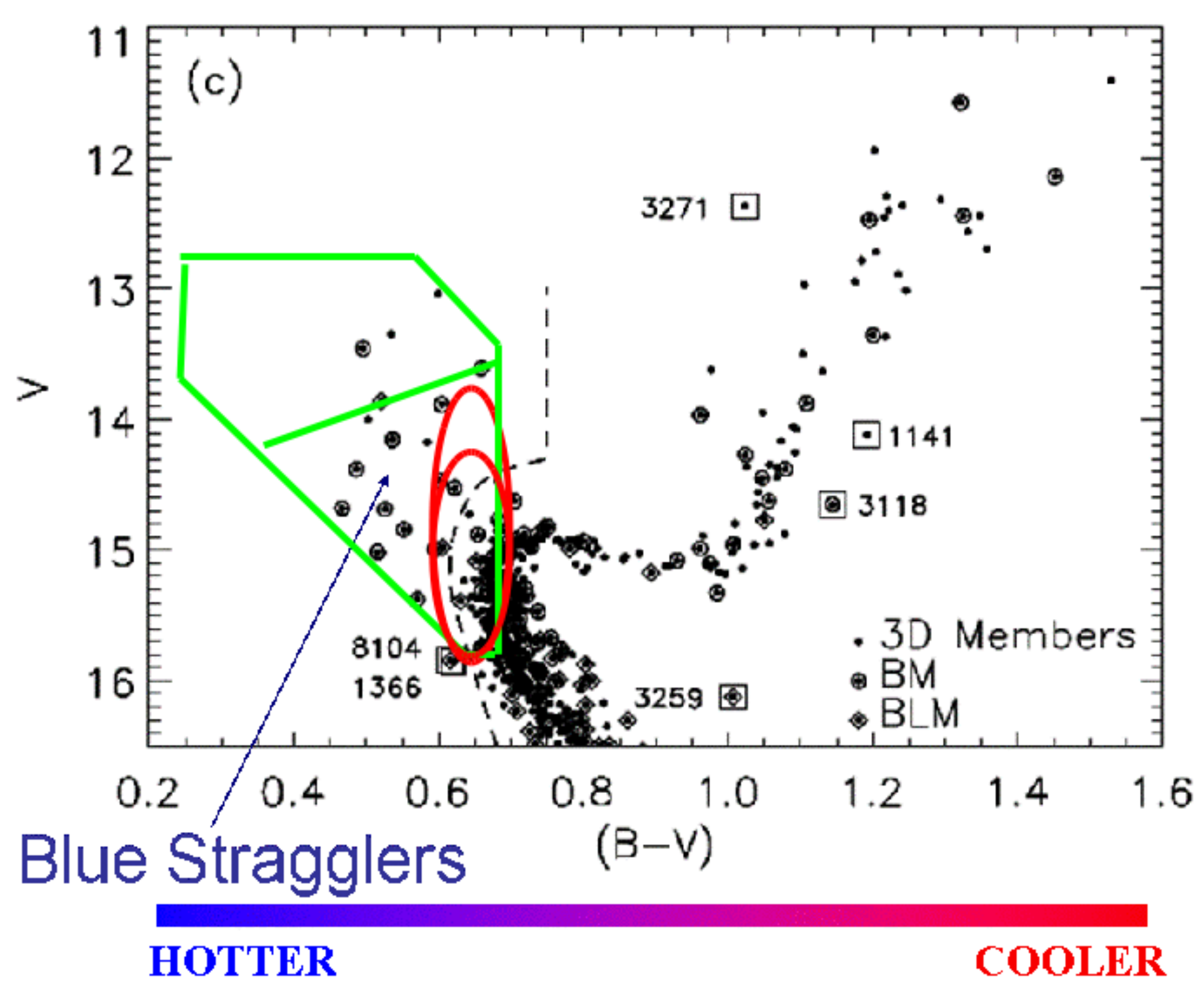}
\caption{Position of NGC 188 BSSs on the CMD. Top left:
Population synthesis results of BSSs formed through case C mass transfer.
Top right: Population synthesis results of BSSs formed through mergers.
Bottom: The location of observed BSSs in the open cluster NGC 188 \cite{gel+08},
compared with the expected location of case C mass transfer from population
synthesis models (red elipses) and merger products from population
synthesis models (green polygons). As can be seen, the observed BSSs
are consistent with being merger products (and since most of them
are in binaries, their progenitors must be triples in this case),
but, for the most, are inconsistent with being case C mass transfer
products. Population synthesis models were made using the BSSE code
(taken from the senior thesis of M. Bailey --- adviser: R. Mathieu). }
\label{fig:CMD-location}
\end{center}
\end{figure}

Figure~\ref{fig:CMD-location} shows the outcome of population synthesis\index{population synthesis}
of evolving binaries resulting in mergers in case A MT, compared with
case C MT. Case C MT can explain the wide orbits of the BSS binaries
in NGC 188\index{NGC 188}, but the resulting BSSs are too faint and cannot explain
the observed BSS population. The CMD\index{colour-magnitude diagram} location of full mergers\index{merger} is consistent
with the observed BSS population, but then a third companion must be
invoked to explain the binarity of the observed BSS populations.

BSS formed through the case D MT could easily have wide orbit companions,
and even their eccentricity needs not affected much. However, the
expected periods for this scenario are higher than the typical periods
observed for BSS binaries in OCs.

%%%%%%%%%%%% FIGURE 4 %%%%%%%%%%%%%%
%%%%%%%%%%%% FIGURE 4 %%%%%%%%%%%%%%

\subsection{Field BSSs}\index{field}
\label{secper:5.3}

\subsubsection{Low Luminosity Halo BSSs}\index{halo}
\label{secper:5.3.1}
Comparing the various expectations and constraints from the currently
suggested models for BSS formation, it appears that low luminosity
field BSSs such as studied by \cite{car+05} could be fully explained
by case B and C MT. Most of these BSSs are found in binary systems
with periods in the range $167-1576$ day period, a typical $\Delta{\rm M}$
in the range $0.03-0.48$~${\rm M_{\odot}}$, with non-circular, but
low eccentricity orbits. The binary companions are not seen, and are
likely WDs; their minimal masses range between $0.18-0.55$~${\rm M_{\odot}}$.
Most of these properties are consistent with the predictions of the
case C MT, beside the BSSs with the largest $\Delta{\rm M}$ which
appear to be too high compared with theoretical predictions. These,
however, might serve as evidence for additional formation route, either
a case B MT or triple evolution scenario. In the case B MT, larger
$\Delta M$ are possible. In this case we expect BSSs with the largest
$\Delta{\rm M}$ to have the lowest mass companions, corresponding
to helium WDs\index{helium white dwarf} expected in the case B MT. We might also expect to generally
find them at shorter periods. One peculiarity is the relatively high
eccentricities\index{eccentricity} of these most massive BSS binaries, which is less clear
in the context of MT. This might be explained by the triple secular
evolution model, in which the BSS binary eccentricities are not expected
to be low, the formed BSS are massive merger products, and the most
typical companions are low mass MS companions. The observed periods,
however, are shorter than seen for the relevant triples in the field
(see Fig. \ref{fig:period-eccentricity}). Unfortunately, the statistics
are currently too few to confirm/refute the eccentricity-mass trend. 

It therefore appears that at least in the case of low luminosity field
BSSs our theoretical understanding of their origin from case C MT with
a contribution from case B and/or triple evolution case is fully consistent
with observations. Taking this path, we can now try and learn about
case B/C MT from these observations. Though the observed orbital eccentricities
are not zero, they are much lower than the typical eccentricities
of binaries with similar periods in the field (with the mentioned
caveat for the massive BSSs), showing that binaries are circularised\index{circularisation},
though at lower rates than expected from current theories. We will
discuss this issue in the context of OC BSSs.

\subsubsection{Other Field BSSs}\index{field}
\label{secper:5.3.2}
Other populations of BSSs exist outside stellar clusters. Their characterisation,
however, is still in an early stage, and will only be briefly mentioned
here. 

\textbf{Luminous halo BSSs} Current studies of halo BSSs
focused on low luminosity BSSs, while luminous, likely more massive
BSSs have hardly been studied. Massive main sequence stars do exist
in the halo\index{halo}, as shown, for example, by the discoveries of hyper-velocity\index{hyper-velocity star}
main sequence B-stars. Most of these stars are thought to be
regular main sequence stars ejected at high velocities from the Galactic
Centre, while a small fraction might have been ejected from dense
young clusters. Some, however, appear too young to have had the time
to propagate from such locations in the Galactic Centre\index{Galactic Centre}, or from young
stellar clusters in the Galactic disc, and might have been rejuvenated
through mass transfer\index{mass transfer}, making them BSSs \cite{per09a}. Additional
observations of halo BSSs are needed in order to characterise their
overall population. 

\textbf{Bulge BSSs} Recently, the first bulge\index{Bulge} BSSs have been
detected: about a quarter of them appears to be in short period W
UMa\index{W UMa star} type binaries \cite{cla+11}. Their overall fraction compared
to background HB population is consistent with that found for halo
BSSs by \cite{car+01}. The low densities in the bulge require the
BSS formation to go through a binary/triple evolution interaction and
not through collisions. From the large number of short period binaries,
it appear that case A MT and mergers are likely the dominant channel.
Given the likely origin of short period binaries in triples \cite{2006EK,2006T,2007FT},
it is suggestive that the dominant route for BSS formation in the bulge
is through triple\index{triple system} evolution leading to mergers and case A MT \cite{per+09}.

\section{Summary}
\label{secper:6}
Blue stragglers exist in a wide variety of environments, ranging from
low density environments such as the Galactic halo\index{halo} and bulge\index{Bulge} through
open clusters\index{open cluster} to dense globular clusters\index{globular cluster}. The amount of observational
data, varies widely from one environment to the other. The largest
sets of data are available for GCs, however the most detailed data
including specific properties of binary BSSs are of open clusters such
as NGC 188\index{NGC 188} and M67\index{M67}, but the latter include only a small number of
BSSs. The study of blue stragglers and their origin touches upon
a wide range of fields, ranging from stellar evolution, stellar collisions,
dynamics of few-body systems and the overall evolution of stellar
clusters. Though the BSS phenomena exist in many environments, it is
not yet clear whether the same processes play similar roles in their
production. Current theories for the origin of BSSs include various
type of mass transfer or merger products in binaries; possible direct
collision origin; or induced merger/collision in secularly evolving
triple systems. 

The advance in complex simulation of clusters which include both dynamics
and stellar evolution provide a wealth of theoretical predictions
which could be compared for the first time with observations. Comparisons
between the simulations and the observations suggest that the data
on BSSs in GCc are consistent with the BSSs having a collisional origin
mostly from binary-single and binary-binary encounters. However, given
the major role of binary/triple systems in the theory of BSS formation
but the lack of information on the binarity of BSSs, it is still premature
to conclude that the BSS formation in GCs is well understood. Much
more detailed data are available for OCs, but current simulations
show striking failure in reproducing the observed populations. The
currently most likely origin of BSSs in OCs is likely a combinations
of induced mergers and case A mass transfer in triples as well as
case C/D mass transfer in evolved binaries. It does appear clear that
collisions play at most a minor role in producing OC BSSs and binary
BSSs in these environments. 

Field BSSs cannot form through collisions due to the low stellar density
environments. It appears that the current data on galactic halo BSSs
are consistent with case B/C mass transfer, though triple secular
evolution may also contribute. Less data is available for bulge BSSs,
but the large fraction of eclipsing binaries among them suggest that
secular triples play an important role (producing the short period
binaries, that then produce BSSs through case A mass transfer and mergers).
Case B/C mass transfer are also likely to play a role in those cases.

We conclude that BSSs are likely to have multiple origins both different
origins in different environments, as well as combinations of various
evolutionary/dynamical channels in the any given environments.

%\newpage
\begin{landscape}
\begin{table}
%\begin{sideways}
\caption{\label{tab:BSS-obs}Summary of BSSs properties in different environments}
%{\tiny
\begin{tabular}{lcccc}
\hline\noalign{\smallskip}
Environment & GC & OC & Halo & Bulge\tabularnewline
Property & \multicolumn{4}{c}{}\tabularnewline
\noalign{\smallskip}\svhline\noalign{\smallskip} 
\textbf{Inferred $\Delta M$ (${\rm M_{\odot})}$ } & $0.2-0.8$ & $0.2-0.8$ & $0.03-0.48$ & \tabularnewline
%\hline 
\textbf{Frequency} & $10^{-5}-10^{-4}$ & $\sim few\times10^{-2}$ & 1/2000 & \tabularnewline
%\hline 
\textbf{Spatial distribution} & Centrally concentrated  & Centrally concentrated  & -- & --\tabularnewline
 & and sometimes bi-model & and sometimes bi-model &  & \tabularnewline
%\hline 
\textbf{Binarity} & Unknown; Large fraction & High; $76$ \% in NGC 188 & High;  & at least 25\% are w UMa \tabularnewline
 & in eclipsing binaries & Consistent with 100\% & consistent with 100\% & short period binaries\tabularnewline
%\hline 
Companion & unknown & Peaks at $0.55$${\rm M_{\odot}}$ & $0.18\le M_{min}\le0.55$ ${\rm M_{\odot}}$ & unknown\tabularnewline
mass &  &  &  & \tabularnewline
%\hline 
Period & unknown & $P>500$ days & $167-1576$ days  & $>25$ \%$P<10$ days \tabularnewline
 &  & $10$\% $P<10$ days & typical $200-800$ days & \tabularnewline
%\hline 
Eccentricity & unknown & High; $<e>\sim0.34$  & Low; $<e>\sim0.17$ & unknown\tabularnewline
 &  &  &  & \tabularnewline
%\hline 
Envrionmental  & $f_{BSS}\propto M_{cluster}$ & too low statistics & -- & --\tabularnewline
correlates & $f_{BSS}\propto f_{bin}$ &  &  & \tabularnewline
%\hline 
\noalign{\smallskip}\hline\noalign{\smallskip}
\end{tabular}
%}
%above is for tiny
%\end{sideways}
\end{table}
%\end{landscape}
%\newpage

%\begin{sidewaystable}
\begin{table}
\caption{\label{tab:BSS-thoery}Summary of BSSs models and their predictions}
{\tiny
\begin{tabular}{ccccccccc}
\hline\noalign{\smallskip}
Theoretical model & MT A / Merger & MT B & MT C & MT D & Collisions & Secular  & GC models  & OC models \tabularnewline
 &  &  &  &  &  & Triples & w/o triples & w/o triples\tabularnewline
Property & \multicolumn{8}{c}{}\tabularnewline
\noalign{\smallskip}\svhline\noalign{\smallskip} 
$\Delta M_{max}$ (${\rm M_{\odot})}$ & turn-off mass & up to a few M$_{\odot}$  & $<0.25$  & $0.5$ turn-off mass & turn-off mass & turn-off mass & turn-off mass & turn-off mass\tabularnewline
 &  & below turn-off mass & typical $<0.15$ &  & and even higher &  &  & \tabularnewline
%\hline 
Composition & regular MS & regular MS & enriched AGB  & enriched AGB  & regular (?) & regular MS & comination of collision/merger & Mostly\tabularnewline
 &  &  & processed elements & processed elements &  &  & and MT products & MT/merger products\tabularnewline
%\hline 
Spatial distribution & -- & -- & -- & -- & Centrally concentrated & -- & Centrally concentrated; & \tabularnewline
 &  &  &  &  &  &  & can be bimodal & \tabularnewline
%\hline 
Binarity & 0 \% (merger) & High 100 \% & High 100 \% & High 100 \% & High (collisions in binaries) & High $100$ \% & $\sim60$\% & Intermediate\tabularnewline
\multirow{1}{*}{} & High 100 \% (MT) &  &  &  &  &  &  & \tabularnewline
%\hline 
Companion & None (merger); & Helium WD & CO WD & CO WD & Any type & Any type & Any type & Mostly CO WDs\tabularnewline
 & MS (MT), likely & ${\rm 0.1\le M}\le0.45$ & ${\rm 0.55\le M\le0.65}$ & ${\rm 0.55\le M\le0.65}$ &  &  &  & \tabularnewline
 & w UMa type binary &  &  &  &  &  &  & \tabularnewline
%\hline 
Period & $P<3-4$ days & typical  & typical  & typical  & typical & $>500$ days; inferred & typical  & Mostly large separations\tabularnewline
 &  & $10<P<1000$ & $300<P<2000$  & $2000<P<10^{4}$ & $10<P<1000$ & from field triples & $10<P<1000$ & $>500$ days\tabularnewline
%\hline 
Eccentricity & Likely circular & Very low & Low & Somewhat low & High & typical outer binaries  & High & Mostly circular\tabularnewline
 &  & circularised & circularised &  &  & of triples$\left\langle e\right\rangle \sim0.3-0.4$ &  & (due to BSSE prescriptions ?)\tabularnewline
%\hline 
\noalign{\smallskip}\hline\noalign{\smallskip}
\end{tabular}
}
%for tiny

%\end{sidewaystable}
\end{table}
\end{landscape}

%\newpage

%

\backmatter%%%%%%%%%%%%%%%%%%%%%%%%%%%%%%%%%%%%%%%%%%%%%%%%%%%%%%%
%\appendix
%\include{appendix}
%\include{glossary}
\printindex

%%%%%%%%%%%%%%%%%%%%%%%%%%%%%%%%%%%%%%%%%%%%%%%%%%%%%%%%%%%%%%%%%%%%%%

\end{document}